\documentclass{article}
\usepackage[T1]{fontenc}
\usepackage{spconf,amsmath,amssymb,graphicx,booktabs,balance}
\usepackage[hidelinks]{hyperref}
\title{CLEAN-ANCHORED EXCESS-EMISSION CONTROL FOR CTC ADAPTATION\\ UNDER COMPETING SPEECH}
\name{Yuanbo Zhou \qquad Junjie Xiong}
\address{Missouri University of Science and Technology\\
\texttt{yztpn@mst.edu, junjiexiong@mst.edu}}
\begin{document}
\raggedbottom
\maketitle

\begin{abstract}
Competing speech can cause excess emissions in a connectionist temporal
classification (CTC) recognizer. We penalize the amount by which the noisy student's expected collapsed token count exceeds a frozen teacher's count on the paired clean utterance. The one-sided objective caches one scalar per utterance and adds no inference module. With 147,456 trainable parameters in wav2vec~2.0, it improves competing-speech word error rate (WER) over matched CTC-only adaptation across three development seeds. On a locked LibriSpeech test, speech-interference WER falls from 45.198\% to 44.323\%, while clean WER changes by 0.004 percentage points. Unseen VCTK interferers yield a 0.307-point improvement. Ablations favor collapsed-count control over frame-mass regularization. Longer training increases the speech gain but worsens noise and music performance, identifying an insertion--deletion tradeoff rather than uniform robustness.
\end{abstract}


\section{Introduction}

Overlapping speech challenges automatic speech recognition (ASR).
Existing approaches use enhanced signals~\cite{sato2021separate},
speaker conditioning~\cite{wang2023target}, or permutation-invariant
training for multiple output streams~\cite{yu2017pit}. We instead
adapt a single-output CTC recognizer to mixtures with a designated
target transcript, without adding an inference-time separator or
speaker enrollment input.

Robustness can be improved through augmentation, including the
spectrogram transformations of SpecAugment~\cite{park2019specaugment},
and teacher-guided learning. Generalized distillation guides a noisy
student using clean-speech teacher outputs~\cite{markov2016distillation};
teacher--student adaptation matches posteriors on parallel
source/target data~\cite{li2017teacherstudent}. Clean teacher guidance
is therefore established prior work. We investigate a different
transferred statistic: the expected collapsed CTC token count.

CTC merges consecutive repeated labels and removes
blanks~\cite{graves2006ctc}. In our development experiments,
CTC-only adaptation under competing speech often produced excess
words. Blank-logit calibration reduced some insertions but did not
eliminate them. We therefore penalize only the amount by which the
noisy student's expected collapsed count exceeds the frozen model's
count on paired clean speech. The auxiliary loss transfers one cached
scalar per utterance; supervised CTC learns token identity. We test
this objective with matched adapters, three development seeds,
component ablations, locked testing, and unseen interfering speakers.

\begin{figure*}[t]
\centering
\includegraphics[width=\textwidth]{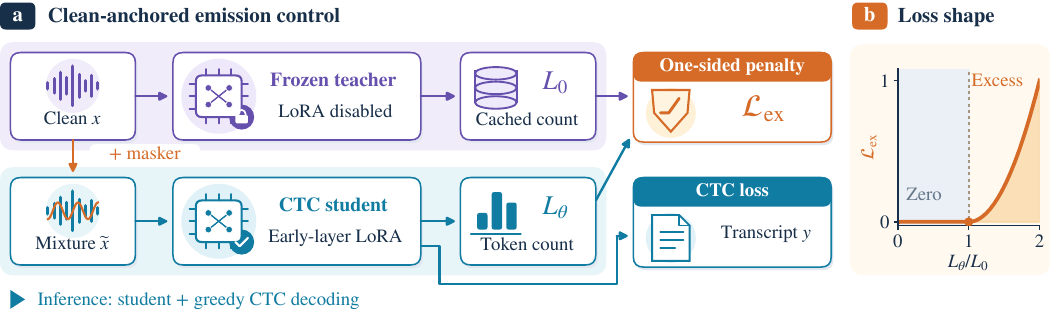}
\caption{Clean-anchored excess-emission control. (a) The frozen
teacher supplies a cached count for each clean utterance. The student
receives the mixture and target transcript; only LoRA weights are
updated. (b) The penalty is zero below the anchor and quadratic above
it ($L_0\geq1$ shown). The teacher is absent at inference.}
\label{fig:method}
\end{figure*}

\subsection{Relation to prior work}

LoRA learns low-rank updates~\cite{hu2022lora}; Deep Filter-Tuning
filters distorted speech representations~\cite{ng2025filter}, and
SSVD-O studies structured speech-model adaptation~\cite{wang2026ssvd}.
We fix adapter placement and rank to isolate the auxiliary objective.

CTC regularizers address different behaviors. Soft forgetting combines
chunked BLSTM training with whole-utterance hidden-state
guidance~\cite{audhkhasi2019softforgetting}; CR-CTC matches predictions
across augmented views~\cite{yao2025crctc}; blank-regularized CTC
encourages blank emissions by restricting consecutive nonblank repetitions, enabling transducer frame skipping~\cite{yang2023blank}.
Our loss instead constrains an utterance-level collapsed count,
without matching full teacher posteriors or rewarding further
shortening below the anchor. It controls output extent, not speaker
identity, and does not guarantee preservation of target tokens.

\section{Clean-Anchored Emission Control}
\subsection{Expected collapsed token count}
Figure~\ref{fig:method} summarizes the training graph. Let $p_t(k)$ be the frame-level softmax probability of token $k$, and let
$\epsilon$ denote the CTC blank. We draw frame labels independently from
$p_t$ and count the nonblank symbols that remain after merging consecutive
repeats and removing blanks. Its expected count has the closed form
\begin{equation}
 \widehat L(p)=\sum_{k\ne\epsilon}p_1(k)+
 \sum_{t=2}^{T}\sum_{k\ne\epsilon}p_t(k)[1-p_{t-1}(k)].
 \label{eq:length}
\end{equation}
To obtain Eq.~\eqref{eq:length}, write the collapsed length as a sum of
indicators. A nonblank label contributes at the first frame; at each later frame it contributes only if it differs from the preceding sampled label. Independence of the frame draws then gives
$\Pr(\pi_t=k,\pi_{t-1}\ne k)=p_t(k)[1-p_{t-1}(k)]$.
Thus the expression is exact under the factorized path distribution. It is a surrogate for decoded length, not an expectation under the transcript-conditioned CTC posterior, and its units are tokens rather than words.

This distinction matters for the alternative frame-mass statistic
$M(p)=\sum_t[1-p_t(\epsilon)]$. For a deterministic path $(a,a,a)$,
$M=3$ but $\widehat L=1$. The path $(a,\epsilon,a)$ yields $M=2$ and
$\widehat L=2$. Frame mass counts occupied frames; Eq.~\eqref{eq:length} counts separate token emissions, including repetitions separated by a blank.

\subsection{One-sided clean anchoring}

For a clean waveform $x$ and corrupted counterpart $\widetilde x$, a frozen teacher gives $L_0=\widehat L(p_0(x))$. The adapted student gives
$L_\theta=\widehat L(p_\theta(\widetilde x))$. We use
\begin{equation}
 \mathcal L_{\rm ex}=
 \left[\max\left(0,\frac{L_\theta-L_0}{\max(L_0,1)}\right)\right]^2,
 \qquad
 \mathcal L=\mathcal L_{\rm CTC}+\lambda\mathcal L_{\rm ex}.
 \label{eq:loss}
\end{equation}
The extra term is applied only to corrupted training examples; clean examples use the ordinary CTC loss. We set $\lambda=1$ before replication. The normalization scales the excess by the clean expected count, with a denominator floor of one. A clean teacher count is cached per utterance. At inference, the teacher and the loss are removed, leaving the same architecture and cost as the CTC-only adapter.

The anchor provides a training signal about output extent, not token identity. The supervised CTC term still learns the target transcript $y$ from each mixture. Once $L_\theta\leq L_0$, the regularizer has zero derivative and leaves the CTC gradient unchanged. This avoids explicitly rewarding ever-shorter
outputs, although it does not guarantee that deletions cannot increase.
An erroneous clean count can also produce an overly restrictive anchor.

\subsection{Adaptation and implementation}
We adapt the query and value projections in encoder layers 0 and 1
of wav2vec~2.0 Base~\cite{baevski2020wav2vec} using rank-24 LoRA.
The backbone and CTC head remain frozen. Each projection uses
$W+BA$, where $A$ has shape $24\times768$ and $B$ has shape
$768\times24$. Both methods update 147,456 parameters across four
projections and differ only in the auxiliary objective.

The count costs $O(T|\mathcal V|)$ elementwise operations on logits already computed for CTC. We evaluate the clean anchor with adapters disabled and the backbone in evaluation mode, then restore student training mode. Caching the scalar avoids a second resident model and repeated clean forwards. The regularizer uses float32 softmax even when the backbone uses mixed precision. No frame alignment between clean and noisy posteriors is required.

\section{Experimental Protocol}
\subsection{Data and training}
We use the public wav2vec2-base-960h checkpoint with greedy decoding. A
deterministic two-hour subset (631 utterances) of Mini LibriSpeech train-clean-5 provides labeled target speech~\cite{panayotov2015librispeech}. Each training draw remains clean with probability 0.5. Otherwise, we mix one recording sampled from a uniformly chosen MUSAN noise, music, or speech family~\cite{snyder2015musan}, at a signal-to-masker energy ratio drawn uniformly from 0 to 20 dB. MUSAN recordings are split 70/15/15 by file for training, development, and test corruption.

Adapters use AdamW, learning rate $10^{-4}$, gradient accumulation 4, and 400
updates, with zero weight decay and gradient clipping at norm 1.
Development uses the 2,703 dev-clean utterances split by speaker into
1,252 calibration and 1,451 validation utterances. Calibration uses the three masker families at 5 and 10 dB. We search a blank
offset $b$ from $-2$ to 2 in 0.25 increments, adding $b$ to the blank logit before greedy decoding. Selection minimizes pooled noisy WER subject to at most 0.2 points of clean degradation relative to the unadapted model at $b=0$. The method and baseline receive the
same search. The locked test uses all 2,620 test-clean utterances at 0, 5, 10, and 20 dB and the seed-29 offset of 0.5.

For external interference, 500 locked test-clean targets are mixed at the same four ratios with 600 utterances from 20 VCTK 0.92 speakers~\cite{yamagishi2019vctk}. These data test transfer of the masker speakers, not transfer of the target speech domain. The test uses seed 29; the three-seed replication is development-only.
Test and external confidence intervals resample the 40 target speakers;
the budget diagnostic contains 20 target speakers. Development-seed variation is reported separately. No language model or speech separator is used.

We micro-average WER from word substitutions, deletions, and insertions. The paired bootstrap draws target speakers with all of their clean and corrupted versions together, using 2,000 replicates. Runs pin the backbone revision and record source, data-manifest, and checkpoint hashes. Every aggregate in this paper was checked against the saved utterance-level predictions.

\subsection{Controls}
The main baseline, CTC-LoRA, uses the same layers, rank, mixtures, optimizer, samples, and decoding calibration. Two seed-29 controls replace our penalty: \emph{symmetric anchor} also penalizes under-emission, while \emph{frame mass} uses $\sum_t(1-p_t(\epsilon))$ instead of Eq.~\eqref{eq:length}. A fixed-bias diagnostic resumes both 400-step checkpoints to 800 updates and evaluates a hash-selected set of 500 validation utterances without reopening selection.

\section{Results}
\subsection{Locked test and replication}
Table~\ref{tab:test} shows the locked test. Excess-emission control reduces speech-interference WER by 0.875 points and pooled noisy WER by 0.341 points. The paired 95\% CI for the pooled change is $[-0.383,-0.298]$ points. Clean WER changes by 0.004 points (CI $[-0.042,0.048]$). Across all noisy conditions, substitutions and insertions fall by 1,534 and 1,258, while deletions rise by 640. For speech interference alone, substitutions, deletions, and insertions change by $-1001$, $-30$, and $-810$.

\begin{table}[t]
\centering
\caption{Locked test WER (\%). $\Delta$ is proposed minus CTC-LoRA in points.}
\label{tab:test}
\small
\setlength{\tabcolsep}{4.1pt}
\begin{tabular}{lrrrrr}
\toprule
Method & Clean & Noise & Music & Speech & Pooled\\
\midrule
CTC-LoRA & 3.492 & \textbf{13.566} & 18.191 & 45.198 & 25.652\\
Proposed & 3.496 & 13.692 & \textbf{17.918} & \textbf{44.323} & \textbf{25.311}\\
$\Delta$ & +.004 & +.126 & -.273 & -.875 & -.341\\
\bottomrule
\end{tabular}
\end{table}

\begin{figure}[t]
\centering
\includegraphics[width=\columnwidth]{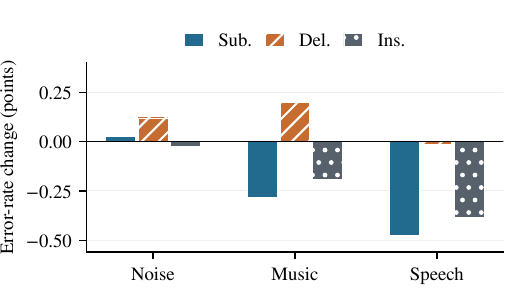}
\caption{Locked-test changes in substitution, deletion, and insertion rates (proposed minus CTC-LoRA). Each family uses the same 210,304 reference words. The speech gain includes fewer substitutions and insertions without increased deletions; noise and music exhibit a deletion tradeoff.}
\label{fig:errors}
\end{figure}

Figure~\ref{fig:errors} shows how the error changes differ by masker. On competing speech, both insertions and substitutions decrease
without a deletion increase. For noise and music, fewer insertions coexist with more deletions. Error counts are consistent with the intended mechanism but do not establish that every removed token originated from the interferer; that would require alignment to the masker's transcript.

The effect replicates on development data. Table~\ref{tab:seeds} reports proposed-minus-baseline changes after the shared calibration protocol. Speech and pooled WER improve for every seed. Environmental noise worsens slightly in all three, while music varies. The sign of the speech gain is stable despite the different training seeds.

\begin{table}[t]
\centering
\caption{Validation WER change (percentage points) over three training seeds.}
\label{tab:seeds}
\small
\setlength{\tabcolsep}{4.3pt}
\begin{tabular}{rrrrrr}
\toprule
Seed & Clean & Noise & Music & Speech & Pooled\\
\midrule
17 & +.044 & +.082 & -.168 & -.467 & -.184\\
29 & +.037 & +.113 & -.093 & -1.055 & -.345\\
43 & -.110 & +.064 & +.084 & -1.128 & -.327\\
Mean & -.010 & +.086 & -.059 & -.884 & -.285\\
\bottomrule
\end{tabular}
\end{table}

\subsection{Which component accounts for the gain?}
Table~\ref{tab:ablation} shows the component controls. Symmetric anchoring loses 0.292 points on speech and 0.067 points pooled relative to the proposed loss. Replacing collapsed count with frame-level nonblank mass removes the gain over CTC-LoRA and yields 324 more insertions than our method. These controls support the specific combination of a one-sided anchor and collapsed count within this configuration. They do not establish that either component is optimal across regularization strengths.

\begin{table}[t]
\centering
\caption{Seed-29 validation ablation at 400 updates. WER and insertion count $I$
are pooled over 5/10 dB unless marked Speech.}
\label{tab:ablation}
\small
\setlength{\tabcolsep}{5pt}
\begin{tabular}{lrrr}
\toprule
Method & Speech WER & Pooled WER & $I$\\
\midrule
CTC-LoRA & 48.486 & 25.341 & 9,430\\
Symmetric anchor & 47.723 & 25.062 & 9,247\\
Frame mass & 48.501 & 25.375 & 9,497\\
Proposed & \textbf{47.431} & \textbf{24.995} & \textbf{9,173}\\
\bottomrule
\end{tabular}
\end{table}

\subsection{Transfer to unseen interfering speakers}
With unseen VCTK interferers, Table~\ref{tab:vctk} shows improvement at every ratio. Overall WER falls from 42.967\% to 42.660\%, or 122 fewer word errors; the paired change is $-0.307$ points with CI $[-0.482,-0.112]$. The gain is modest but extends beyond the MUSAN speech recordings used during development.

\begin{table}[t]
\centering
\caption{WER (\%) with unseen VCTK interfering speakers.}
\label{tab:vctk}
\small
\setlength{\tabcolsep}{5.4pt}
\begin{tabular}{lrrrrr}
\toprule
Method & 0 dB & 5 dB & 10 dB & 20 dB & All\\
\midrule
CTC-LoRA & 72.693 & 53.909 & 34.742 & 10.524 & 42.967\\
Proposed & \textbf{72.532} & \textbf{53.406} & \textbf{34.390} & \textbf{10.313} & \textbf{42.660}\\
\bottomrule
\end{tabular}
\end{table}

\subsection{Training budget and specialization}
\begin{figure}[t]
\centering
\includegraphics[width=\columnwidth]{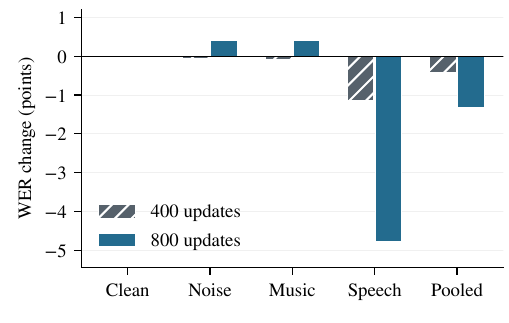}
\caption{Training-budget sensitivity on the same 500 development utterances
(seed 29, fixed blank offset 0.5). Bars show proposed minus CTC-LoRA WER; negative is better. Speech gains grow with training, while noise and music regress. These are two measured endpoints, not a learning curve.}
\label{fig:budget}
\end{figure}

Both methods benefit from extending training to 800 updates. The relevant comparison is the gap at an equal training budget. On the fixed 500-utterance subset, pooled WER improves from 24.801\% to 22.122\% for CTC-LoRA and from 24.375\% to 20.800\% for the proposed method. The proposed-minus-baseline gap grows from $-0.426$ to $-1.322$ points (CI $[-1.636,-1.024]$ at 800). Its speech gap grows from $-1.128$ to $-4.782$ points (Fig.~\ref{fig:budget}). However, the 800-update method is worse than the baseline on noise and music by 0.414 and 0.403 percentage points.

The 800-update speech improvement comprises 428 fewer substitutions and 517 fewer insertions, offset by 55 additional deletions. For noise and music, deletions instead rise by 75 and 108, exceeding the insertion reductions. Longer training shifts the observed tradeoff toward fewer insertions and more deletions. These counts do not identify which token-level gradients cause the change. The diagnostic uses development data only and does not replace the locked 400-update test results. The 12.1\% relative speech-WER reduction on this subset cannot substitute for the 1.94\% reduction measured on the locked 400-update test.

\subsection{Scope of the evidence}
The baseline comparison isolates the auxiliary objective; it does not compare against other adaptation or target-speaker methods. The VCTK experiment retains LibriSpeech targets and therefore does not test transfer to a new target domain.

The error decomposition does not establish a causal gradient-conflict
mechanism. Neither the clean anchor nor the one-sided loss guarantees
preservation of target tokens, and the regularization strength has not been optimized independently for each masker.

The paired confidence intervals capture target-speaker variation at fixed
checkpoints and corruptions, excluding uncertainty from retraining or
masker resampling. The three development seeds assess training variability
separately. Pooled WER depends on the evaluated balance of masker families;
per-family results remain necessary when that balance differs in deployment.

Our test mixtures cover target-to-masker ratios from 0 to 20 dB,
so the target has at least as much overall energy as the masker.
Evaluating negative ratios would help distinguish robustness in
these conditions from gains confined to mixtures with an
equal-energy or stronger target.

\section{Conclusion}
Clean-anchored control of expected collapsed emissions improves early-layer CTC adaptation when the masker is another speaker, with no inference-time module. Locked and external gains are modest; environmental noise can regress. The evidence covers one backbone, synthetic mixtures, and greedy decoding; it does not establish target-speaker identification or robustness to natural overlap. Future evaluation should test those settings and whether the clean anchor remains reliable on less accurate backbones.

\clearpage
\section{Compliance with Ethical Standards}
This study uses publicly available LibriSpeech, MUSAN, and
VCTK recordings. No participants were recruited and no new
recordings were collected for this study.


\interlinepenalty=10000
\balance
\bibliographystyle{IEEEbib}
\bibliography{references}
\end{document}